\documentclass[prd,twocolumn,showpacs,superscriptaddress,floatfix]{revtex4}

\usepackage{ifthen}
\usepackage{lineno}
\usepackage{epsfig}
\usepackage{amsmath}
\usepackage{comment}
\usepackage{longtable}
\usepackage{subfigure}
\usepackage{color}
\usepackage{xspace}

\usepackage{multirow}
\usepackage{array}
\usepackage{booktabs}
\usepackage{dcolumn}
\usepackage[abs]{overpic}

\newcommand{\vmet}{\vec{{\mbox{$\protect \raisebox{0.3ex}{$\not$}E$}}}_{T}}
\newcommand{\vmpt}{{\mbox{$\protect \raisebox{0.3ex}{$\not$}\vec{p}$}}_{T}}

\newcommand{\Et}{\mbox{$E_T$}}
\newcommand{\Pt}{\mbox{$p_T$}}
\newcommand{\Ht}{\mbox{$H_T$}}

\newcommand{\met}{\mbox{$\protect \raisebox{0.3ex}{$\not$}\Et$}}
\newcommand{\mpt}{\mbox{$\protect \raisebox{0.3ex}{$\not$}\Pt$}}
\newcommand{\mht}{\mbox{$\protect \raisebox{0.3ex}{$\not$}\Ht$}}
\newcommand{\gevcc}{\mbox{GeV/$c^2$}}
\newcommand{\mjj}{\mbox{$m_{\it jj}$}}

\begin{document}

\title{Search for a dijet resonance in events with jets and missing transverse energy in 
\boldmath{$p\bar{p}$} collisions at \boldmath{$\sqrt{s} = 1.96$} TeV}

\affiliation{Institute of Physics, Academia Sinica, Taipei, Taiwan 11529, Republic of China}
\affiliation{Argonne National Laboratory, Argonne, Illinois 60439, USA}
\affiliation{University of Athens, 157 71 Athens, Greece}
\affiliation{Institut de Fisica d'Altes Energies, ICREA, Universitat Autonoma de Barcelona, E-08193, Bellaterra (Barcelona), Spain}
\affiliation{Baylor University, Waco, Texas 76798, USA}
\affiliation{Istituto Nazionale di Fisica Nucleare Bologna, \ensuremath{^{ii}}University of Bologna, I-40127 Bologna, Italy}
\affiliation{University of California, Davis, Davis, California 95616, USA}
\affiliation{University of California, Los Angeles, Los Angeles, California 90024, USA}
\affiliation{Instituto de Fisica de Cantabria, CSIC-University of Cantabria, 39005 Santander, Spain}
\affiliation{Carnegie Mellon University, Pittsburgh, Pennsylvania 15213, USA}
\affiliation{Enrico Fermi Institute, University of Chicago, Chicago, Illinois 60637, USA}
\affiliation{Comenius University, 842 48 Bratislava, Slovakia; Institute of Experimental Physics, 040 01 Kosice, Slovakia}
\affiliation{Joint Institute for Nuclear Research, RU-141980 Dubna, Russia}
\affiliation{Duke University, Durham, North Carolina 27708, USA}
\affiliation{Fermi National Accelerator Laboratory, Batavia, Illinois 60510, USA}
\affiliation{University of Florida, Gainesville, Florida 32611, USA}
\affiliation{Laboratori Nazionali di Frascati, Istituto Nazionale di Fisica Nucleare, I-00044 Frascati, Italy}
\affiliation{University of Geneva, CH-1211 Geneva 4, Switzerland}
\affiliation{Glasgow University, Glasgow G12 8QQ, United Kingdom}
\affiliation{Harvard University, Cambridge, Massachusetts 02138, USA}
\affiliation{Division of High Energy Physics, Department of Physics, University of Helsinki, FIN-00014, Helsinki, Finland; Helsinki Institute of Physics, FIN-00014, Helsinki, Finland}
\affiliation{University of Illinois, Urbana, Illinois 61801, USA}
\affiliation{The Johns Hopkins University, Baltimore, Maryland 21218, USA}
\affiliation{Institut f\"{u}r Experimentelle Kernphysik, Karlsruhe Institute of Technology, D-76131 Karlsruhe, Germany}
\affiliation{Center for High Energy Physics: Kyungpook National University, Daegu 702-701, Korea; Seoul National University, Seoul 151-742, Korea; Sungkyunkwan University, Suwon 440-746, Korea; Korea Institute of Science and Technology Information, Daejeon 305-806, Korea; Chonnam National University, Gwangju 500-757, Korea; Chonbuk National University, Jeonju 561-756, Korea; Ewha Womans University, Seoul, 120-750, Korea}
\affiliation{Ernest Orlando Lawrence Berkeley National Laboratory, Berkeley, California 94720, USA}
\affiliation{University of Liverpool, Liverpool L69 7ZE, United Kingdom}
\affiliation{University College London, London WC1E 6BT, United Kingdom}
\affiliation{Centro de Investigaciones Energeticas Medioambientales y Tecnologicas, E-28040 Madrid, Spain}
\affiliation{Massachusetts Institute of Technology, Cambridge, Massachusetts 02139, USA}
\affiliation{University of Michigan, Ann Arbor, Michigan 48109, USA}
\affiliation{Michigan State University, East Lansing, Michigan 48824, USA}
\affiliation{Institution for Theoretical and Experimental Physics, ITEP, Moscow 117259, Russia}
\affiliation{University of New Mexico, Albuquerque, New Mexico 87131, USA}
\affiliation{The Ohio State University, Columbus, Ohio 43210, USA}
\affiliation{Okayama University, Okayama 700-8530, Japan}
\affiliation{Osaka City University, Osaka 558-8585, Japan}
\affiliation{University of Oxford, Oxford OX1 3RH, United Kingdom}
\affiliation{Istituto Nazionale di Fisica Nucleare, Sezione di Padova, \ensuremath{^{jj}}University of Padova, I-35131 Padova, Italy}
\affiliation{University of Pennsylvania, Philadelphia, Pennsylvania 19104, USA}
\affiliation{Istituto Nazionale di Fisica Nucleare Pisa, \ensuremath{^{kk}}University of Pisa, \ensuremath{^{ll}}University of Siena, \ensuremath{^{mm}}Scuola Normale Superiore, I-56127 Pisa, Italy, \ensuremath{^{nn}}INFN Pavia, I-27100 Pavia, Italy, \ensuremath{^{oo}}University of Pavia, I-27100 Pavia, Italy}
\affiliation{University of Pittsburgh, Pittsburgh, Pennsylvania 15260, USA}
\affiliation{Purdue University, West Lafayette, Indiana 47907, USA}
\affiliation{University of Rochester, Rochester, New York 14627, USA}
\affiliation{The Rockefeller University, New York, New York 10065, USA}
\affiliation{Istituto Nazionale di Fisica Nucleare, Sezione di Roma 1, \ensuremath{^{pp}}Sapienza Universit\`{a} di Roma, I-00185 Roma, Italy}
\affiliation{Mitchell Institute for Fundamental Physics and Astronomy, Texas A\&M University, College Station, Texas 77843, USA}
\affiliation{Istituto Nazionale di Fisica Nucleare Trieste, \ensuremath{^{qq}}Gruppo Collegato di Udine, \ensuremath{^{rr}}University of Udine, I-33100 Udine, Italy, \ensuremath{^{ss}}University of Trieste, I-34127 Trieste, Italy}
\affiliation{University of Tsukuba, Tsukuba, Ibaraki 305, Japan}
\affiliation{Tufts University, Medford, Massachusetts 02155, USA}
\affiliation{University of Virginia, Charlottesville, Virginia 22906, USA}
\affiliation{Waseda University, Tokyo 169, Japan}
\affiliation{Wayne State University, Detroit, Michigan 48201, USA}
\affiliation{University of Wisconsin, Madison, Wisconsin 53706, USA}
\affiliation{Yale University, New Haven, Connecticut 06520, USA}

\author{T.~Aaltonen}
\affiliation{Division of High Energy Physics, Department of Physics, University of Helsinki, FIN-00014, Helsinki, Finland; Helsinki Institute of Physics, FIN-00014, Helsinki, Finland}
\author{S.~Amerio\ensuremath{^{jj}}}
\affiliation{Istituto Nazionale di Fisica Nucleare, Sezione di Padova, \ensuremath{^{jj}}University of Padova, I-35131 Padova, Italy}
\author{D.~Amidei}
\affiliation{University of Michigan, Ann Arbor, Michigan 48109, USA}
\author{A.~Anastassov\ensuremath{^{v}}}
\affiliation{Fermi National Accelerator Laboratory, Batavia, Illinois 60510, USA}
\author{A.~Annovi}
\affiliation{Laboratori Nazionali di Frascati, Istituto Nazionale di Fisica Nucleare, I-00044 Frascati, Italy}
\author{J.~Antos}
\affiliation{Comenius University, 842 48 Bratislava, Slovakia; Institute of Experimental Physics, 040 01 Kosice, Slovakia}
\author{G.~Apollinari}
\affiliation{Fermi National Accelerator Laboratory, Batavia, Illinois 60510, USA}
\author{J.A.~Appel}
\affiliation{Fermi National Accelerator Laboratory, Batavia, Illinois 60510, USA}
\author{T.~Arisawa}
\affiliation{Waseda University, Tokyo 169, Japan}
\author{A.~Artikov}
\affiliation{Joint Institute for Nuclear Research, RU-141980 Dubna, Russia}
% ...

\collaboration{CDF Collaboration}
\altaffiliation[With visitors from]{
\ensuremath{^{a}}University of British Columbia, Vancouver, BC V6T 1Z1, Canada,
\ensuremath{^{b}}Istituto Nazionale di Fisica Nucleare, Sezione di Cagliari, 09042 Monserrato (Cagliari), Italy,
\ensuremath{^{c}}University of California Irvine, Irvine, CA 92697, USA,
\ensuremath{^{d}}Institute of Physics, Academy of Sciences of the Czech Republic, 182~21, Czech Republic,
\ensuremath{^{e}}CERN, CH-1211 Geneva, Switzerland,
\ensuremath{^{f}}Cornell University, Ithaca, NY 14853, USA,
\ensuremath{^{g}}University of Cyprus, Nicosia CY-1678, Cyprus,
\ensuremath{^{h}}Office of Science, U.S. Department of Energy, Washington, DC 20585, USA,
\ensuremath{^{i}}University College Dublin, Dublin 4, Ireland,
\ensuremath{^{j}}ETH, 8092 Z\"{u}rich, Switzerland,
\ensuremath{^{k}}University of Fukui, Fukui City, Fukui Prefecture, Japan 910-0017,
\ensuremath{^{l}}Universidad Iberoamericana, Lomas de Santa Fe, M\'{e}xico, C.P. 01219, Distrito Federal,
\ensuremath{^{m}}University of Iowa, Iowa City, IA 52242, USA,
\ensuremath{^{n}}Kinki University, Higashi-Osaka City, Japan 577-8502,
\ensuremath{^{o}}Kansas State University, Manhattan, KS 66506, USA,
\ensuremath{^{p}}Brookhaven National Laboratory, Upton, NY 11973, USA,
\ensuremath{^{q}}Queen Mary, University of London, London, E1 4NS, United Kingdom,
\ensuremath{^{r}}University of Melbourne, Victoria 3010, Australia,
\ensuremath{^{s}}Muons, Inc., Batavia, IL 60510, USA,
\ensuremath{^{t}}Nagasaki Institute of Applied Science, Nagasaki 851-0193, Japan,
\ensuremath{^{u}}National Research Nuclear University, Moscow 115409, Russia,
\ensuremath{^{v}}Northwestern University, Evanston, IL 60208, USA,
\ensuremath{^{w}}University of Notre Dame, Notre Dame, IN 46556, USA,
\ensuremath{^{x}}Universidad de Oviedo, E-33007 Oviedo, Spain,
\ensuremath{^{y}}CNRS-IN2P3, Paris, F-75205 France,
\ensuremath{^{z}}Universidad Tecnica Federico Santa Maria, 110v Valparaiso, Chile,
\ensuremath{^{aa}}The University of Jordan, Amman 11942, Jordan,
\ensuremath{^{bb}}Universite catholique de Louvain, 1348 Louvain-La-Neuve, Belgium,
\ensuremath{^{cc}}University of Z\"{u}rich, 8006 Z\"{u}rich, Switzerland,
\ensuremath{^{dd}}Massachusetts General Hospital, Boston, MA 02114 USA,
\ensuremath{^{ee}}Harvard Medical School, Boston, MA 02114 USA,
\ensuremath{^{ff}}Hampton University, Hampton, VA 23668, USA,
\ensuremath{^{gg}}Los Alamos National Laboratory, Los Alamos, NM 87544, USA,
\ensuremath{^{hh}}Universit\`{a} degli Studi di Napoli Federico I, I-80138 Napoli, Italy
}
\noaffiliation

\date{\today}

\begin{abstract}
We report on a search for a dijet resonance in events with only two or three jets and 
large imbalance in the total event transverse momentum.  This search is sensitive to 
the possible production of a new particle in association with a $W$ or $Z$ boson, 
where the boson decays leptonically with one or more neutrinos in the final state. 
We use the full data set collected by the CDF~II detector at the Tevatron collider 
at a proton-antiproton center-of-mass energy of 1.96 TeV. These data correspond to 
an integrated luminosity of 9.1 fb$^{-1}$. We study the invariant mass distribution 
of the two jets with highest transverse energy. We find good agreement between data 
and standard model background expectations and measure the combined cross section 
for {\it WW}, {\it WZ}, and {\it ZZ} production to be $13.8^{+3.0}_{-2.7}$ pb. 
No significant anomalies are observed in the mass spectrum and 95\% credibility 
level upper limits are set on the production rates of a potential new particle in 
association with a $W$ or $Z$ boson.
\end{abstract}

\pacs{12.15.Ji, 12.38.Qk, 14.70.--e, 14.80.--j}

\maketitle
\section{\label{sec:Intro}Introduction}

A study of the dijet invariant mass (\mjj) distribution in events with jet pairs produced 
in association with a $W$ boson was performed by the CDF collaboration using $p\bar{p}$ 
collision data corresponding to an integrated luminosity of 4.3 fb$^{-1}$~\cite{ACC-prl}. 
That analysis focused on $W$ boson decays to $\ell\nu$ ($\ell = e$ or $\mu$), where the 
presence of an identified electron ($e$) or muon ($\mu$) was required in the event selection. 
Reference~\cite{ACC-prl} reported evidence for a discrepancy with the standard model (SM) 
expectations, interpretable as an excess of events in the mass range of 120--160 \gevcc\ 
corresponding to a significance of 3.2 standard deviations. In that study, the excess 
could be modeled with a Gaussian distribution, centered at 145 \gevcc\ with an rms width 
of 14.3 \gevcc, corresponding to the expected experimental $m_{\it jj}$ resolution of the 
CDF~II detector. The acceptance and selection efficiencies for events associated with such 
a dijet resonance were estimated by simulating Higgs boson ($H$) production in association 
with a $W$ boson for a Higgs boson mass of $150$ \gevcc. Based on the assumption that the 
observed excess originated from a hypothetical new particle $X$ with a branching fraction 
to quark pairs of one, the excess corresponded to a measured production cross section for 
$\sigma(p\bar{p}\rightarrow {\it WX})$ of $3.1\pm 0.8$ pb~\cite{wjj_xsec}.

In this article, we present a search for a dijet resonance produced in association with a 
vector boson by studying the $\mjj$ distribution from the two highest energy clusters of 
particles (jets) in events with only two or three detected jets and large imbalance in 
total event transverse-momentum, indicative of the presence of undetected particles. We 
veto events containing one or more identified high-$p_T$ leptons, in order to ensure that 
the sample is statistically independent from those used in other studies. The resulting 
final states are sensitive to {\it WX} $\rightarrow {\mbox{\protect $\not{\ell}$}} \nu jj$ 
and {\it ZX} $\rightarrow \nu\bar{\nu} jj$ production and decay, where {\mbox{\protect 
$\not{\ell}$}} represents a hadronically-decaying $\tau$ lepton or an unidentified $e$ or 
$\mu$. We use the entire CDF $p\bar{p}$ collision data set corresponding to an integrated 
luminosity of 9.1 fb$^{-1}$. 

The production of both {\it WX} and {\it ZX} states is of interest since many of the 
theoretical models proposed to explain the excess at 145 \gevcc\ allow the hypothetical 
particle $X$ to be produced in association with either a $W$ boson or a $Z$ boson. 
While studies on {\it WX} production are presented in Refs.~\cite{ACC-prl, d0-mjj-prl, 
ref:cms-dijet}, no studies focusing on {\it ZX} production have been reported to date. 
The search for {\it WX} and {\it ZX} production in events with jets and an imbalance of 
transverse energy is analogous to the search for {\it WH} and {\it ZH} production in 
the same final state~\cite{cdf-metbb}, which has comparable sensitivity to that for the 
{\it WH} process reconstructed in the final state with a lepton and jets, but is based 
on an independent event sample.

\section{\label{sec:Data}Data Sample and Event Preselection}

The data were collected by CDF II\,\cite{cdf-detector}, a general-purpose detector used 
to study Tevatron $p\bar{p}$ collisions at a center-of-mass energy of 1.96 TeV. CDF~II 
features a charged-particle tracking system consisting of a cylindrical open-cell drift 
chamber and silicon microstrip detectors immersed in a 1.4 T magnetic field parallel to 
the beam axis.  Electromagnetic and hadronic calorimeters surrounding the tracking system 
measure the energies of charged and neutral particles. Drift chambers and scintillators 
located outside the calorimeter identify muons.

The calorimeter system consists of lead-scintillator sampling electromagnetic and 
iron-scintillator sampling hadronic calorimeters. The calorimeters comprise central 
barrel ($|\eta| \leq 1.1$) and plug ($1.1 \leq |\eta| \leq 3.6$) sections in 
pseudorapidity ($\eta$) space~\cite{ref:pseudo}. Calorimeter modules are arranged in 
a projective-tower geometry. Individual towers in the central barrel subtend 0.1 in 
$|\eta|$ and 15$^{\circ}$ in $\phi$~\cite{ref:pseudo}. The sizes of the towers in the 
end plug calorimeter vary with $|\eta|$, subtending 0.1 in $|\eta|$ and 7.5$^{\circ}$ 
in $\phi$ at $|\eta| = 1.1$, and 0.5 in $|\eta|$ and 15$^{\circ}$ in $\phi$ at 
$|\eta|=3.6$.

Jets are reconstructed from energy deposits in contiguous groups of calorimeter towers, 
using the {\sc jetclu} clustering algorithm~\cite{ref:jetclu} with a fixed cone size 
of $\Delta R \equiv \sqrt{(\Delta\eta)^2 + (\Delta\phi)^2}$ = 0.4. Jet energies are 
corrected~\cite{ref:jetcorr} for nonuniformities of the calorimeter response as a 
function of $\eta$, energy contributions from multiple $p\bar{p}$ interactions 
within the event, and the nonlinear response of the calorimeters.  In contrast with 
the analysis described in Ref.~\cite{ACC-prl}, additional corrections are applied to 
the reconstructed jets in simulated events to more accurately model the energy scales 
of particle showers initiated by quarks and gluons.  These corrections are obtained 
by comparing predicted and observed distributions of the transverse energy balance, 
$p_T(Z/\gamma) - E_T(\mathrm{jet})$, from independent $Z$+1 jet and $\gamma$+1 jet 
event samples~\cite{ref:qgjes_wes}.

We consider events selected online due to the presence of large missing transverse 
energy~\cite{ref:met-def}.  We inclusively select events with $\met > 45$~GeV and
also the additional events with $\met > 30$~GeV that contain two reconstructed jets.
The event missing transverse energy is corrected offline for the presence of muons,
which typically deposit only a fraction of their energy in the calorimeter, and 
reconstructed charged-particle tracks pointing at inactive regions of the detector.
In order to only retain events for which the online selection is fully efficient, 
we require those selected for further analysis to have a corrected $\met > 50$~GeV. 

We additionally require events to contain two or three reconstructed jets, where 
the two with the highest transverse energies~\cite{ref:jet-et}, $j_1$ and $j_2$,
meet minimal threshold requirements of $E_T(j_1)>35$~GeV and $E_T(j_2)>25$~GeV.
Both jets are required to be reconstructed within the range $|\eta(j_i)|<2$ and 
at least one of the two within $|\eta(j_i)|<0.9$.  We also require the two jets 
to be separated by $\Delta R(j_1,j_2) > 1$.  Events containing only one additional
jet with $E_T>15$~GeV and $|\eta|<2.4$ are not rejected, in order to increase 
acceptance for signal events with an extra jet originating from an initial- or 
final-state radiation or a hadronically-decaying $\tau$ lepton in the final state.  
Events containing an identified electron or muon with $p_T > 20$~GeV/$c$ are 
rejected to maintain orthogonality with other search samples.  Those events that 
satisfy all of the above criteria form the preselection sample used for this 
analysis.  

\section{\label{sec:Bak}Background Modeling}

We model SM background processes using a variety of Monte Carlo (MC) simulation 
programs.  The diboson processes ({\it WW}, {\it WZ}, and {\it ZZ}) are generated 
with {\sc pythia}\,\cite{ref:pythia} incorporating $\gamma^{*}$ contributions to 
the $Z$ boson components for masses above 2 \gevcc. The normalization of simulated 
samples is extrapolated from next-to-leading order calculations~\cite{ref:XsecDIB1, 
ref:XsecDIB2} with the $\gamma^{*}$ and $Z$ contributions restricted to the mass range 
between 40 and 140 \gevcc, yielding cross sections of 11.7 pb for {\it WW}, 3.6 pb 
for {\it WZ}, and 1.5 pb for {\it ZZ} processes, respectively. Top-quark production is 
generated assuming a top-quark mass of 172.5 GeV/$c^2$~\cite{ref:top-mass}. Top-quark 
pair production is generated with {\sc pythia}, and its contribution is normalized 
to the approximate next-to-next-to-leading order cross section~\cite{ref:topxs}. 
Single top-quark production is modeled using {\sc powheg}~\cite{ref:powheg} and 
normalized to the next-to-leading order cross sections~\cite{ref:XsecST1,ref:XsecST2}. 
Production of a $W$ or $Z$ boson in association with parton jets is modeled by 
{\sc alpgen}~\cite{ref:alpgen} incorporating {\sc pythia} to simulate parton 
showering and hadronization. Normalizations for predicted event rates associated 
with these processes are obtained from data. 

\begin{figure}[!htb]
\begin{overpic}[width=0.49\textwidth]{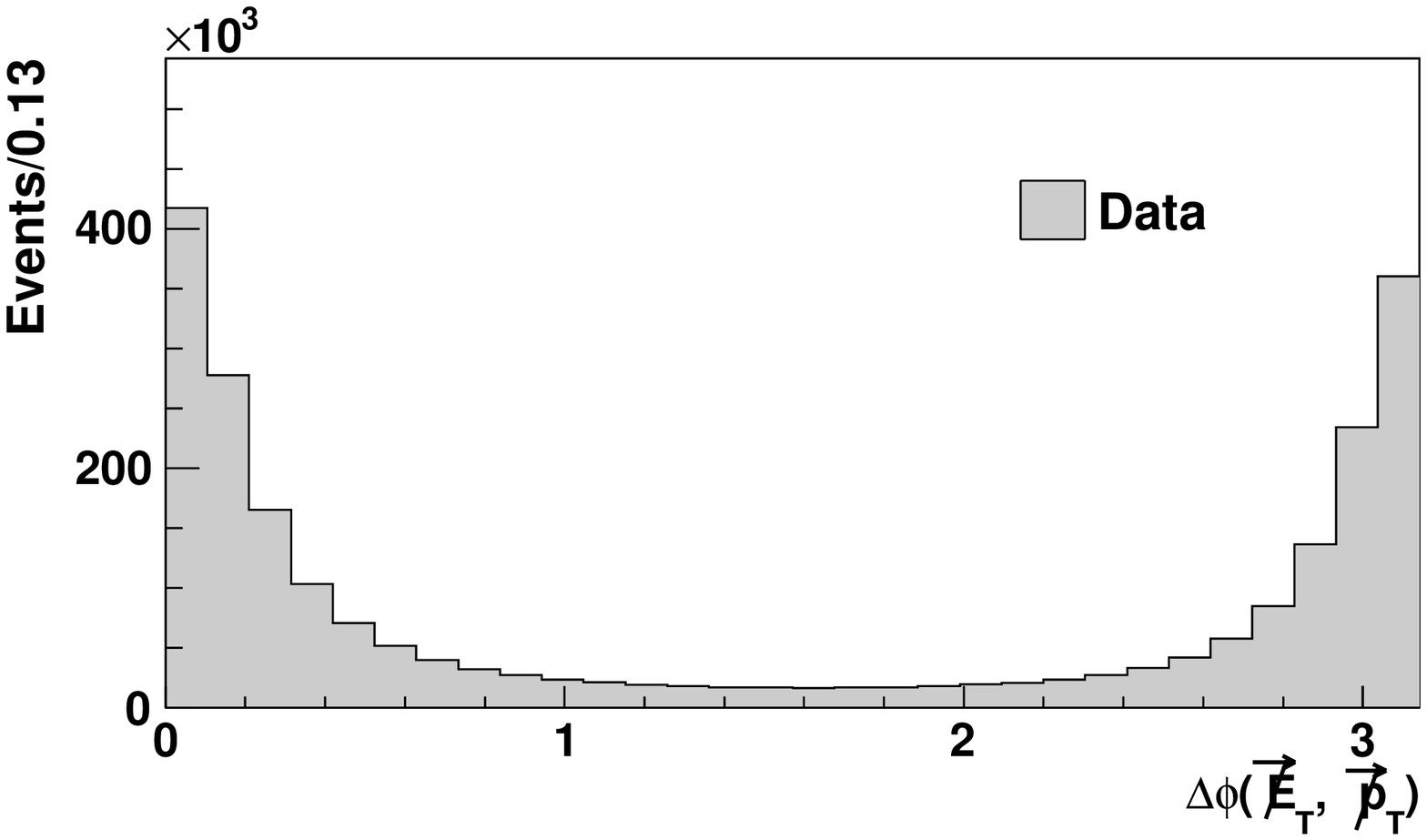}
\put(200,120){\bf(a)}
\end{overpic}
\begin{overpic}[width=0.49\textwidth]{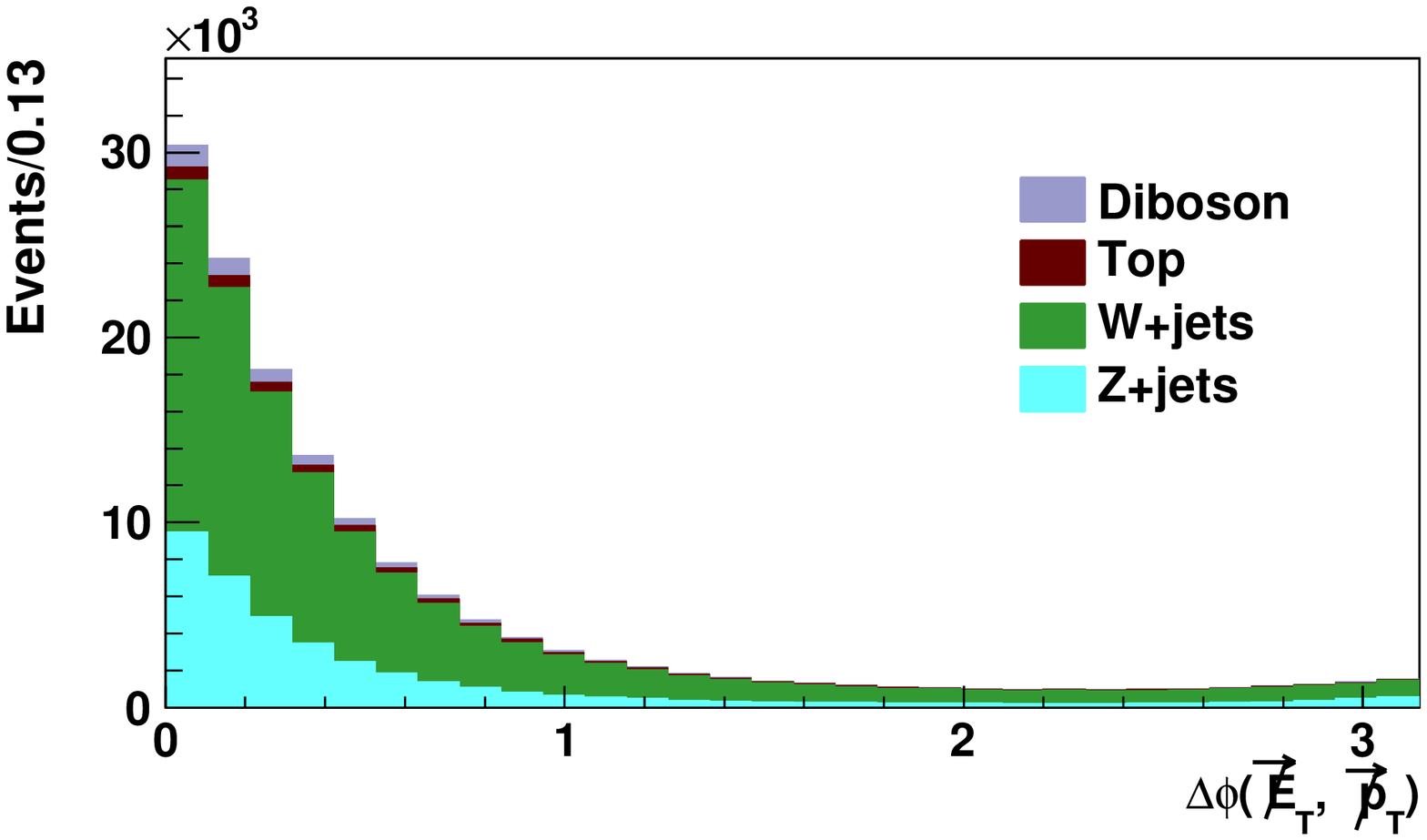}
\put(200,120){\bf(b)}
\end{overpic}
\caption{Azimuthal separation between the $\vmet$ and $\vmpt$ for events that satisfy 
the preselection requirements.  (a) The data distribution, for which 94\% of events 
are estimated to originate from multijet production with observed $\vmet$ and $\vmpt$ 
that tend to be either aligned or antialigned.  (b) Modeled distributions for the 
contributing SM processes leading to events containing final state neutrinos with 
observed $\vmet$ and $\vmpt$ that tend to be aligned.}
\label{ref:dphi-met-mpt}
\end{figure}

We model multijet events from Quantum Chromodynamics processes, a major source of 
background in final states with jets and $\met$, using a data-driven method. We 
define the missing transverse momentum $\vmpt$, a variable similar to $\vmet$, as the 
negative vector sum of charged-particle transverse momenta from the reconstructed 
tracks in an event. As shown in Fig.~\ref{ref:dphi-met-mpt}, $\vmet$ and $\vmpt$ 
tend to be aligned for processes with neutrinos in the final state, such as 
diboson production, but aligned or antialigned in the data, which are dominated 
by multijet production.  Because multijet processes result in final states with no 
neutrinos, observed $\vmet$ necessarily originates from jet energy mismeasurements 
and therefore tends to point either in the same direction or direction opposite to 
the reconstructed $\vec{E}_T$ of the mismeasured jet.  Conversely, observed $\vmpt$ 
in these events is generated from differences in the fractions of showering particles 
within each jet that are reconstructable as charged tracks, a mechanism uncorrelated 
with calorimeter energy mismesurements. Hence, the directions of the observed $\vmet$ 
and $\vmpt$ in these events are in many cases different from one another.  For events 
originating from dijet production, in which the two jets are produced opposite to one 
another, the azimuthal separation between the $\vmet$ and $\vmpt$ thus peaks in the 
regions near 0 or $\pi$.  Hence, multijet background can be suppressed by rejecting 
events where $\Delta\phi(\vmet,\vmpt)>\pi/2$, and rejected events can be used to 
model the multijet background contained within the selected data sample defined 
by $\Delta\phi(\vmet,\vmpt)<\pi/2$. The applicability of this model is confirmed 
in data control regions~\cite{ref:mpt-paper} and supported by other 
measurements~\cite{cdf-metbb, ref:single-top, ref:tprime}.

\section{Analysis Method}

Event preselection yields over 2 million candidate events, of which 94\% 
are estimated to originate from multijet production.  Requiring $\Delta\phi
(\vmet,\vmpt)<\pi/2$, reduces the multijet contribution by roughly a factor 
of two.  To further reduce this background contribution, we require the 
azimuthal separation between the $\vmet$ and each jet to satisfy  $\Delta
\phi(\vmet,j_i)> 0.8$. We also require $\mpt > 20$~GeV and large $\met$ 
significance ($\met/\sqrt{\sum E_T}> 3.5$ GeV$^{1/2}$, where $\sum E_T$ is 
the scalar sum of transverse energies deposited in the calorimeter), as well 
as $\mht/\met<1.2$, where $\mht$ is the magnitude of the negative vector sum 
of jet transverse energies. These additional selections reduce the multijet 
background by more than 99\% and increase $S/\sqrt{B}$ to 11.7 from 3.3, 
where $S$ is the predicted number of SM diboson events and $B$ is the 
predicted number of events from other SM processes in the selected samples.

To study the features of the $\mjj$ distribution in the final event sample, 
we fit the observed distribution in data to the modeled distributions for 
the contributing background processes.  Any contribution from {\it WX} 
and {\it ZX} production would appear as an additional narrow structure 
overlapping the expected, resonant contribution from SM diboson production. 
First, we extract a measurement of diboson production by fitting the $\mjj$ 
distribution for the relative event contributions from known SM processes 
and compare the result with theoretical predictions. We then allow for an 
additional Gaussian contribution from {\it WX} and {\it ZX} production and 
set 95\% credibility level (C.L.) upper limits on the cross section for such 
processes using various theoretical constructs. The fits used to extract 
cross sections and upper limits are based on the Bayesian marginal likelihood 
method~\cite{ref:tom-CL}.

\begin{figure}[!htb]
\includegraphics[width=0.49\textwidth]{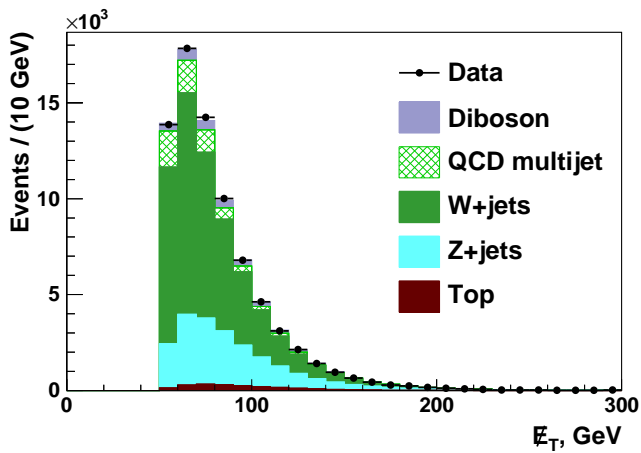}
\caption{Missing transverse energy distribution of events satisfying all selection 
criteria with fitted SM contributions overlaid.  The last bin includes overflow 
events with $\met > 300$~GeV.}
\label{ref:met}
\end{figure}

\begin{figure}[!htb]
\begin{overpic}[width=0.49\textwidth]{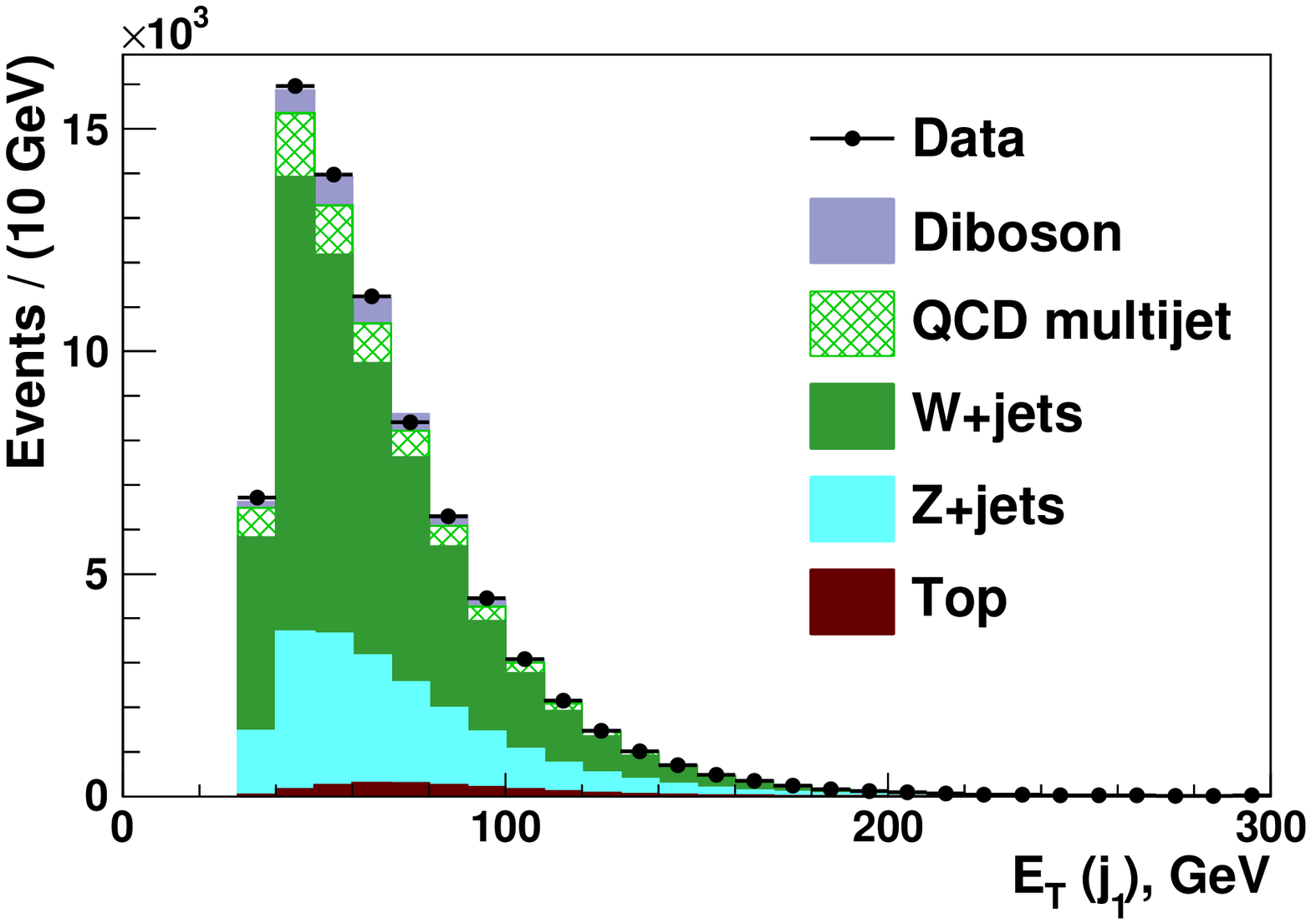}
\put(200,145){\bf(a)}
\end{overpic}
\begin{overpic}[width=0.49\textwidth]{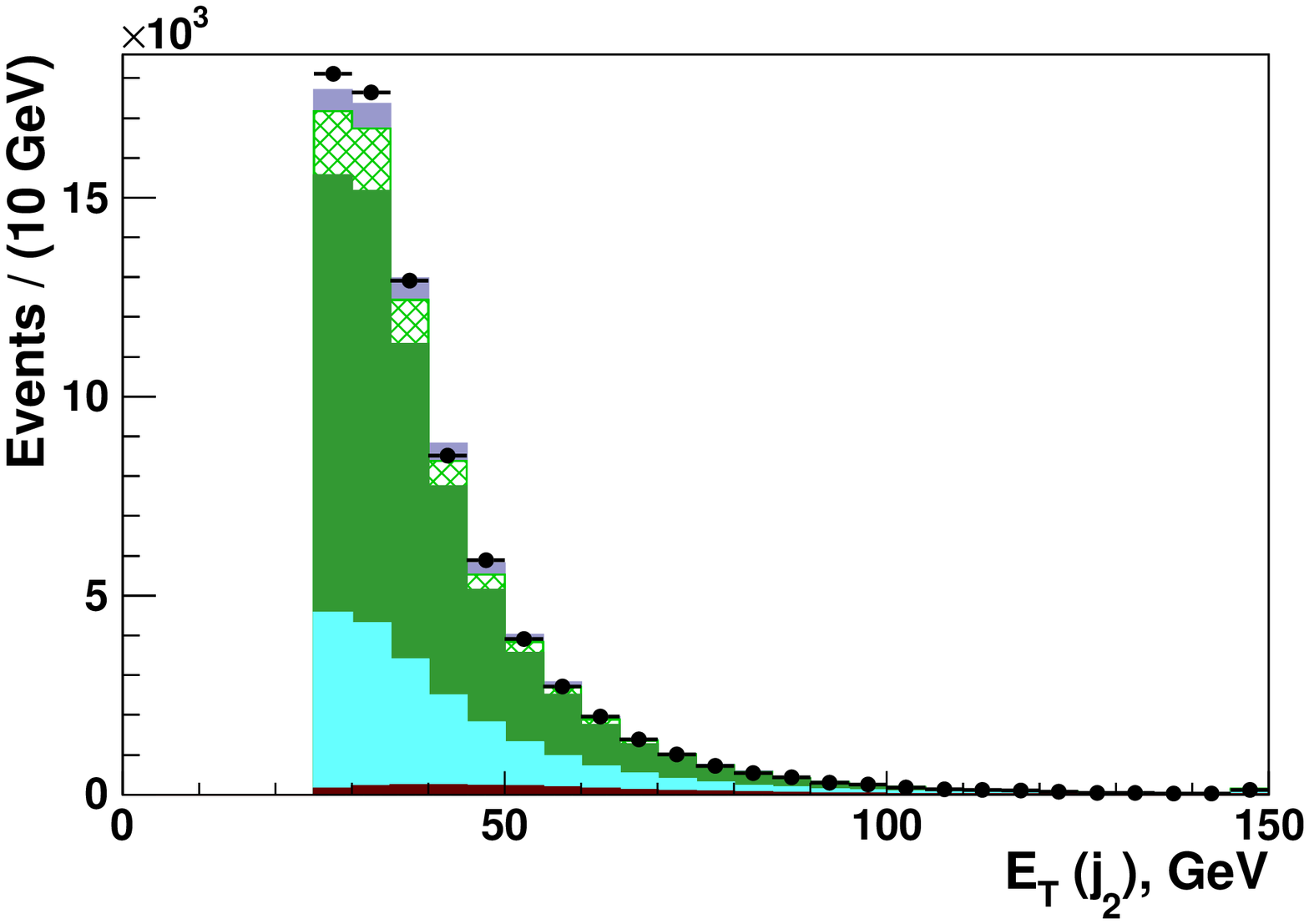}
\put(200,145){\bf(b)}
\end{overpic}
\begin{overpic}[width=0.49\textwidth]{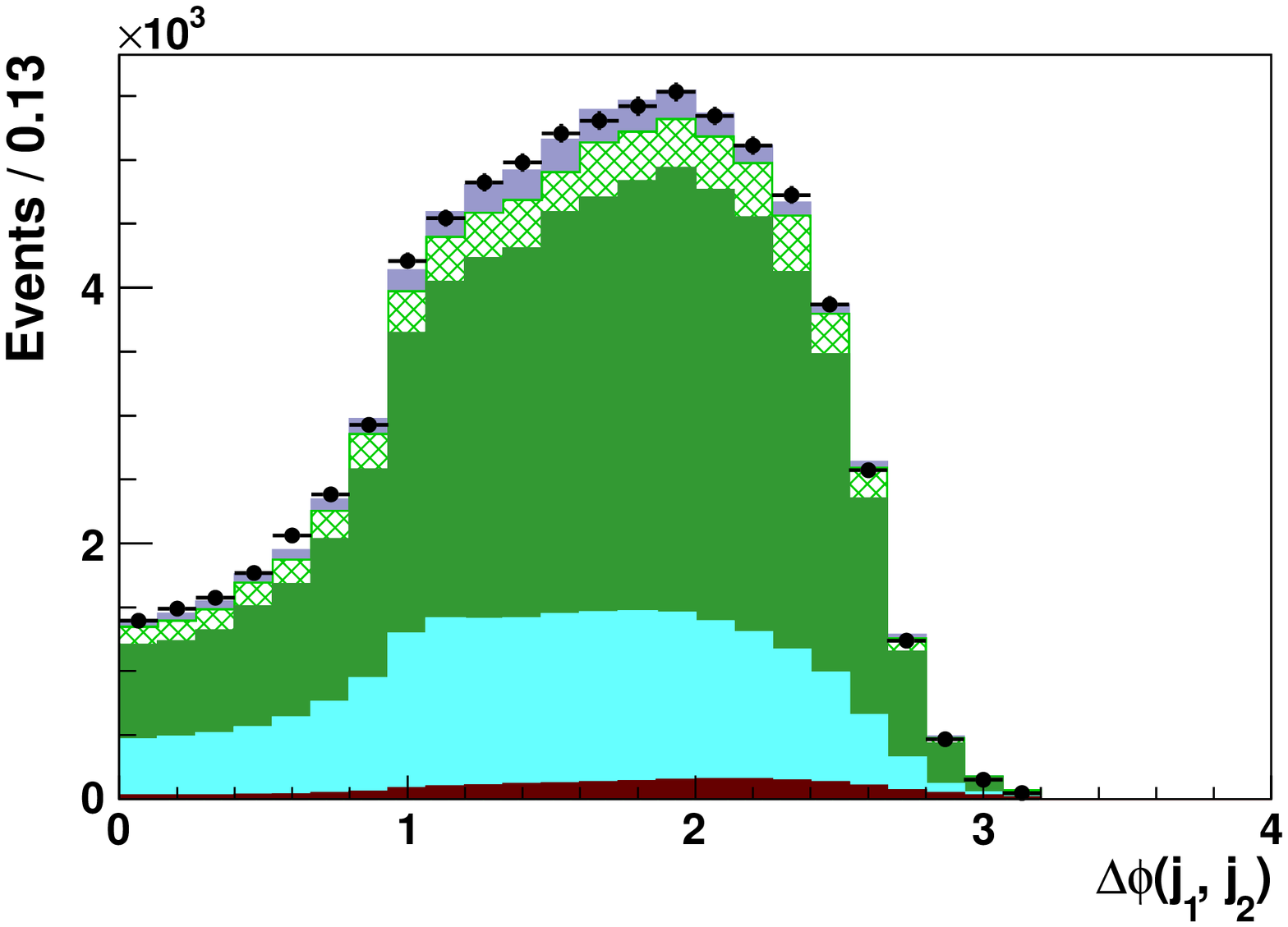}
\put(200,145){\bf(c)}
\end{overpic}
\caption{Distributions of (a) $E_T(j_1)$, (b) $E_T(j_2)$,
and (c) $\Delta\phi(j_1, j_2)$ for events satisfying all 
selection criteria with fitted SM contributions overlaid.
The last bins of the distributions shown in (a) and (b) 
include overflow events with $E_T(j_1) > 300$~GeV and 
$E_T(j_2) > 150$~GeV, respectively.}
\label{ref:jet-et}
\end{figure}

In the final fits, contributions from top-quark production are constrained based 
on theoretical predictions. Initial normalizations for the $W/Z$+jets and multijet 
background contributions are obtained by fitting the $\met$ distribution, which 
provides good discrimination between signal-like and background-like processes, 
using a $\chi^2$ minimization technique. Figure~\ref{ref:met} shows the fitted 
$\met$ distribution, where the $W/Z$+jets and multijet contributions are initially 
treated as unconstrained and determined from the fit.  The resulting uncertainties 
on the multijet and $W/Z$+jets contributions originating from this procedure are 
19\% and 3\%, respectively.  Table~\ref{ref:yield} summarizes predicted event 
contributions to the final event sample from diboson production and other SM 
background processes, which are taken as inputs to the final fits performed on the 
observed $\mjj$ spectrum.  Figure~\ref{ref:jet-et} shows comparisons of predicted 
and observed distributions for $E_T(j_1)$, $E_T(j_2)$, and $\Delta\phi(j_1, j_2)$, 
variables strongly correlated with dijet invariant mass, from events in the final 
sample. 

\begin{table}[!htb]
\begin{center}
\caption{Predicted number of events from each contributing SM process in the final 
event sample and the total number of observed events, where the normalization of 
$W/Z$+jets and multijet background processes are obtained from a fit to the $\met$ 
distribution. Uncertainties include statistical and systematic contributions.}
\label{ref:yield}
\begin{ruledtabular}
\begin{tabular}{lc}
Process &  Yield\\
\hline
{\it WW} &   1850 $\pm$    170\\
{\it WZ} &    670 $\pm$    60\\
{\it ZZ} &    380 $\pm$    30\\
\hline
Top quark & 2040 $\pm$ 190\\
$W$+jets & 46170 $\pm$ 1390\\
$Z$+jets & 19710 $\pm$ 590\\
multijet & 6280 $\pm$ 1190\\
\hline
Total expected & 77100 $\pm$  2320\\
\hline
Data & 77149 \\
\end{tabular}
\end{ruledtabular}
\end{center}
\end{table}

When performing the maximum likelihood fits, we consider several sources 
of systematic uncertainties, included as constraints in the likelihood. 
Sources that affect predicted event yields for modeled background 
contributions are referred to as rate uncertainties.  Dominant rate 
uncertainties include those on the normalizations obtained from data 
to constrain multijet (19\%) and $W/Z$+jets (3\%) contributions.  
Uncertainties associated with theoretical cross section calculations 
(6--7\%) and the sample luminosity measurement~\cite{ref:Lumi} (6\%),
which affect predicted background process event rates taken directly 
from simulation, are also included.  In addition, uncertainty sources 
such as jet energy scale~\cite{ref:jetcorr} (1.4--13\%), parton 
density functions (2\%), efficiency of lepton veto requirements (2\%), 
and measured trigger efficiencies (0.4--1.5\%) that affect simulated 
detector event acceptances are incorporated on both the signal and 
background contributions.    

We also incorporate the effects of systematic uncertainty sources, which 
result in variations in the shapes of modeled $m_{\it jj}$ distributions 
for the contributing processes. For those processes modeled via simulation, 
we account for potential variations in the shape of the $m_{\it jj}$ 
distribution originating from jet energy scale uncertainties.  Uncorrelated 
uncertainties on the simulated energy scales for jets originating from 
quarks (3\%) and gluons (6\%) are considered separately.  In the case of 
the $W/Z$+jets background contribution, shape uncertainties resulting from 
factor of two changes to the nominal Q$^2$ scale used in the perturbative 
expansion for calculating matrix elements in the {\sc alpgen} generator
are also incorporated.  Finally, for the modeled $m_{\it jj}$ distribution 
from multijet production, we obtain shape uncertainties by varying the 
normalization of the modeled contributions from other processes, which 
are subtracted from the data distribution obtained from events with 
$\Delta\phi(\met,\mpt) > \pi/2$.

\section{\label{sec:results}Results}
\subsection{Diboson Measurement}

\begin{figure}[!htb]
\includegraphics[width=0.49\textwidth]{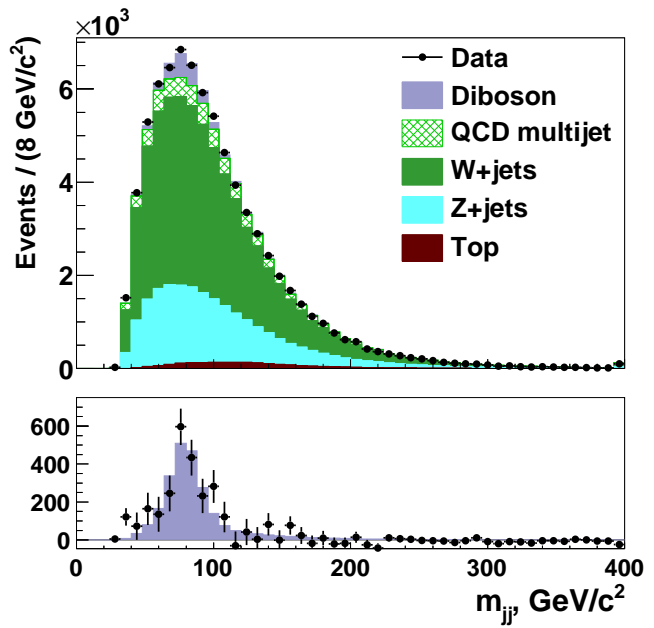}
\caption{Dijet invariant mass distribution with fit results overlaid for 
events passing all selection criteria (top) and the same background-subtracted 
distribution with the fitted diboson contribution overlaid (bottom).   The 
last bin includes overflow events with $\mjj > 400$~GeV/$c^2$.}
\label{ref:mjj} 
\end{figure}

We fit the distribution of $\mjj$ from the two highest-energy jets in events
passing all selection criteria to extract a cross section measurement for 
diboson production.   We take SM values for the relative production rates of 
the {\it WW}, {\it WZ}, and {\it ZZ} processes in order to obtain a single 
$\mjj$ template corresponding to combined diboson production.  The diboson 
contribution is allowed to float freely by assuming a flat, non-negative 
prior probability for the total cross section.  The unit Gaussian priors of 
the nuisance parameters are centered on zero and truncated whenever the value 
results in a nonphysical prediction. Figure~\ref{ref:mjj} shows the fitted 
$m_{\it jj}$ distribution and a comparison of the fitted diboson contribution 
against the data after subtracting the other background contributions obtained 
from the fit.  The inclusive cross section $\sigma(p\bar{p}\rightarrow 
{\it VV})$, where ${\it VV} = {\it WW} + {\it WZ} + {\it ZZ}$, is measured to 
be $13.8^{+ 3.0}_{- 2.7}$ pb for $\gamma^{*}$ and $Z$ contributions restricted 
to the mass range between 40 and 140 \gevcc, which is in good agreement with 
the SM prediction of $16.8\pm1.0$ pb.

\begin{figure}
\includegraphics[width=0.49\textwidth]{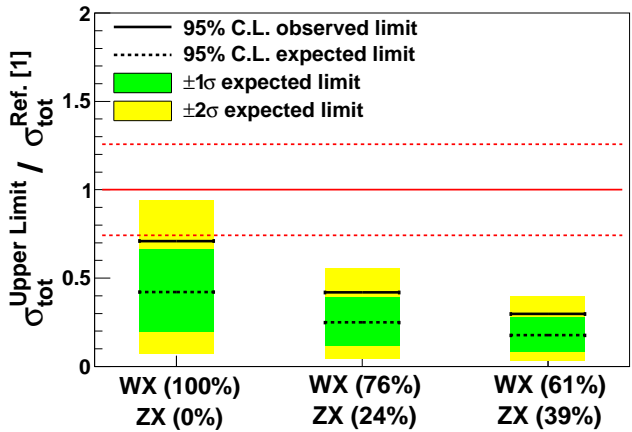}
\caption{\label{fig:yield}Observed and median expected, assuming the background
only hypothesis, 95\% C.L. upper limits on the combined {\it WX}+{\it ZX} cross
section ($\sigma_{\mathrm{tot}}$) divided by the most likely value extrapolated 
from the excess reported in Ref.~\cite{ACC-prl} for each of the three signal 
scenarios.  The measurement uncertainty associated with the {\it WX} cross 
section~\cite{wjj_xsec} assigned to the excess described in Ref.~\cite{ACC-prl} 
is indicated by the light (red) dashed lines.}
\end{figure}

\begin{figure}[!htb]
\includegraphics[width=0.49\textwidth]{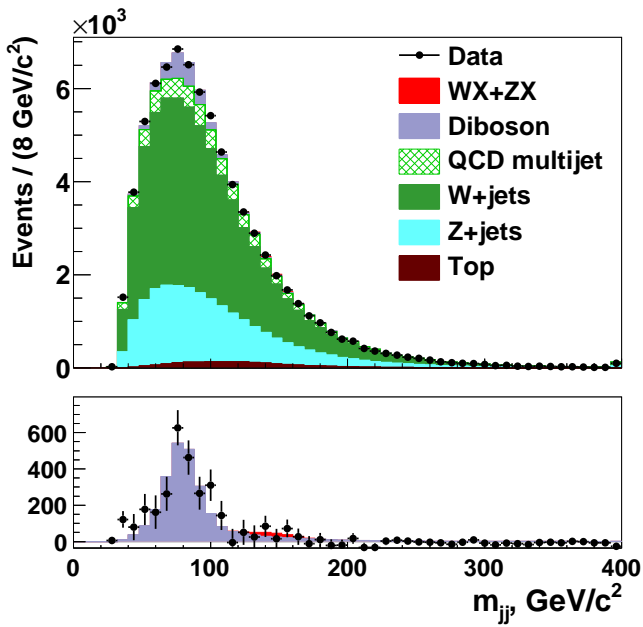}
\caption{Dijet invariant mass distribution with fit results overlaid for events 
passing all selection criteria (top) and the same background-subtracted distribution 
with the fitted contributions from diboson and combined {\it WX}/{\it ZX} production 
overlaid (bottom).   The last bin includes overflow events with $\mjj > 400$~GeV/$c^2$.}
\label{ref:mjj_gaus}
\end{figure}

\subsection{Limits on Dijet-Resonance Cross Sections}

To search for {\it WX} and {\it ZX} production, we perform a second fit, normalizing 
diboson contributions to their theoretical expectation and assuming 6\% uncertainties 
on their theoretical cross sections. We allow for an additional signal contribution, 
modeled assuming a Gaussian distribution, centered at 145 \gevcc\ with an rms width 
of 14.3 \gevcc, in accordance with Ref.~\cite{ACC-prl}.  To be consistent with the
cross section reported in Ref.~\cite{wjj_xsec}, we model the signal acceptance from 
simulated Higgs boson production in association with a $W$ or $Z$ boson for a Higgs 
boson mass of 150 \gevcc\ to extract cross section limits.  As the relative production 
rate of {\it WX} and {\it ZX} varies among theoretical models, we set upper limits 
on combined production ($\sigma_{\mathrm{tot}} = \sigma_{\it WX} + \sigma_{\it ZX}$) 
for three scenarios: (1) $\sigma_{\it WX} = $ 100\% and $\sigma_{\it ZX} =$ 0\%, (2) 
$\sigma_{\it WX} =$ 76\% and $\sigma_{\it ZX} =$ 24\%, and (3) $\sigma_{\it WX} =$ 
61\% and $\sigma_{\it ZX} =$ 39\% .  The second and third scenarios correspond 
approximately to the relative SM rates for {\it WZ}/{\it ZZ} and {\it WH}/{\it ZH} 
production, respectively.  

\begin{table}[!ht]
\caption{\label{tab:limit}Median expected, assuming the background only hypothesis, 
and observed 95\% C.L. upper limits on the combined {\it WX}+{\it ZX} cross section 
($\sigma_{\mathrm{tot}}$) under various hypotheses for the relative magnitudes of 
$\sigma_{\it WX}$ and $\sigma_{\it ZX}$.}
\begin{ruledtabular}
\begin{tabular}{ccc}
\multirow{3}{*}{Signal scenario} & Expected & Observed \\
	&  upper limit &  upper limit \\  & on $\sigma_{\mathrm{tot}}$ & on $\sigma_{\mathrm{tot}}$ \\ 
\hline
$\sigma_{\it WX}/\sigma_{\mathrm{tot}}=$1.00, $\sigma_{\it ZX}/\sigma_{\mathrm{tot}}=$0.00 & 1.31 pb & 2.20 pb \\
$\sigma_{\it WX}/\sigma_{\mathrm{tot}}=$0.76, $\sigma_{\it ZX}/\sigma_{\mathrm{tot}}=$0.24 & 1.02 pb & 1.72 pb \\
$\sigma_{\it WX}/\sigma_{\mathrm{tot}}=$0.61, $\sigma_{\it ZX}/\sigma_{\mathrm{tot}}=$0.39 & 0.90 pb & 1.52 pb \\
\end{tabular}
\end{ruledtabular}
\end{table}

The observed and median expected, assuming the background only hypothesis, 95\% C.L. 
upper limits on the combined cross section for {\it WX} and {\it ZX} production
($\sigma_{\mathrm{tot}}$) obtained from the fit are shown in Table~\ref{tab:limit} 
for each of the three scenarios.  Figure~\ref{fig:yield} shows a comparison of these 
limits relative to expectations for each scenario based on the {\it WX} production 
cross section extrapolated from the observed excess reported in Ref.~\cite{ACC-prl}.
The one standard deviation uncertainties associated with this measured cross section 
are indicated by the light (red) dashed lines.  In all scenarios, the most likely 
value of the combined {\it WX} and {\it ZX} production cross section corresponding 
to the observed excess is excluded at the 95\% C.L.  Figure~\ref{ref:mjj_gaus} 
shows the fitted $m_{\it jj}$ distribution and a comparison of the combined fitted 
contributions of dibosons and {\it WX/ZX} production against the data, with other 
background contributions as obtained from the fit subtracted.

\section{Conclusion}

We study the dijet invariant mass distribution in events with energetic jets and 
large missing transverse energy using the full CDF~II data set corresponding to 
an integrated luminosity of 9.1 fb$^{-1}$.  A fit of the observed distribution to 
modeled distributions for the expected contributions of SM production processes 
gives a measured cross section for combined diboson production of $\sigma(p\bar{p}
\rightarrow {\it VV}) = 13.8^{+ 3.0}_{- 2.7}$ pb, in good agreement with the SM 
prediction of $16.8\pm1.0$ pb.  In the absence of a significant deviation from 
the background expectation in the dijet invariant mass spectrum, we set 95\% C.L. 
upper limits on the combined cross section for the production of a new particle $X$ 
in association with a $W$ or $Z$ boson, under several hypotheses for the relative 
production rates of {\it WX} and {\it ZX}. For each of these hypotheses, we exclude 
at 95\% C.L. the most likely value of the combined {\it WX}+{\it ZX} cross section 
corresponding to the excess observed in Ref.~\cite{ACC-prl}.

\section{Acknowledgments}

We thank the Fermilab staff and the technical staffs of the 
participating institutions for their vital contributions. 
This work was supported by the U.S. Department of Energy and 
National Science Foundation; the Italian Istituto Nazionale di 
Fisica Nucleare; the Ministry of Education, Culture, Sports, 
Science and Technology of Japan; the Natural Sciences and
Engineering Research Council of Canada; the National Science 
Council of the Republic of China; the Swiss National Science 
Foundation; the A.P. Sloan Foundation; the Bundesministerium 
f\"ur Bildung und Forschung, Germany; the Korean World Class 
University Program, the National Research Foundation of Korea; 
the Science and Technology Facilities Council and the Royal 
Society, UK; the Russian Foundation for Basic Research; 
the Ministerio de Ciencia e Innovaci\'{o}n, and Programa
Consolider-Ingenio 2010, Spain; the Slovak R\&D Agency; 
the Academy of Finland; the Australian Research Council (ARC); 
and the EU community Marie Curie Fellowship contract 302103.


\begin{thebibliography}{99}
  \bibitem{ACC-prl}
  T.~Aaltonen {\sl et al.}, (CDF Collaboration),
  Phys. Rev. Lett. {\bf 106}, 171801 (2011).

  \bibitem{wjj_xsec}
  V.~Cavaliere,
  PoS EPS-HEP2011 (2011) 215. 

  \bibitem{d0-mjj-prl}
  V.M. Abazov {\sl et al.}, (D0 Collaboration),
  Phys. Rev. Lett. {\bf 107}, 011804 (2011).
  
  \bibitem{ref:cms-dijet}
  S. Chatrchyan {\it et al.} (CMS Collaboration),
  Phys. Rev. Lett. {\bf 109}, 251801 (2012).
  
  \bibitem{cdf-metbb}
  T. Aaltonen {\sl et al.}, (CDF Collaboration),
  Phys. Rev. Lett. {\bf109}, 111805 (2012).
  
  \bibitem{ref:lvbb-higgs}
  T. Aaltonen {\sl et al.}, (CDF Collaboration),
  Phys. Rev. Lett. {\bf109}, 111804 (2012).

  \bibitem{cdf-detector}
  D. Acosta {\sl et al.}, (CDF Collaboration),
  Phys. Rev. D {\bf 71}, 032001 (2005).

  \bibitem{ref:pseudo} CDF uses a cylindrical coordinate system with the {\it z} axis along 
  the proton beam axis. Pseudorapidity is $\eta = -\mathrm{ln}(\mathrm{tan}( \theta /2))$, 
  where $\theta$ is the polar angle relative to the proton beam direction, and $\phi$ is the 
  azimuthal angle while $p_T$ and $E_T$ are defined as $p_T=|p| \sin \theta$, $E_T=E \sin 
  \theta$.

  \bibitem{ref:jetclu} 
  F.~Abe {\it et al.} (CDF collaboration), 
  Phys Rev. D {\bf 45}, 001448 (1992).
  
  \bibitem{ref:jetcorr} A.~Bhatti {\it et al.}, 
  Nucl. Instrum. Methods A {\bf 566}, 375 (2006).

  \bibitem{ref:qgjes_wes} 
  T.~Aaltonen {\sl et al.}, (CDF Collaboration),
  submitted to Phys. Rev. D, arXiv:1310.0086 (2013).  
  
  \bibitem{ref:met-def} Missing transverse energy, $\met$ , is defined as the magnitude of 
  the vector $\vmet = - \sum_i E_T^i \vec{n}_i$ where $E_T^i$ is the magnitude of transverse 
  energy contained in each calorimeter tower $i$, and $\vec{n}_i$ is the unit vector from 
  the interaction vertex to the tower in the transverse $(x,y)$ plane.

  \bibitem{ref:jet-et} Jet transverse energy, $E_T$, is defined as $\sum^{N}_{i=0} E_i 
  \sin \theta_{jet}$ where $E_i$ is the energy deposited in each of $N$ calorimeter towers, 
  and $\theta$ is the polar angle of the reconstructed jet cluster.

  \bibitem{ref:pythia} 
  T.~Sjostrand, S.~Mrenna, and P.~Skands, 
  J. High Energy Phys. 05 (2006) 026.

  \bibitem{ref:XsecDIB1} 
  J.~M.~Campbell and R.~K.~Ellis, 
  Phys. Rev. D {\bf 60}, 113006 (1999).

  \bibitem{ref:XsecDIB2} 
  J.~M.~Campbell and R.~K.~Ellis, 
  Phys. Rev. D {\bf 62}, 114012 (2000).

  \bibitem{ref:top-mass}
  T.~Aaltonen {\it et al.}  (CDF and D0 Collaborations),
  Phys.\ Rev.\ D {\bf 86}, 092003 (2012).

  \bibitem{ref:topxs} 
  U.~Langenfeld, S.~Moch, and P.~Uwer, 
  Phys. Rev. D {\bf 80}, 054009 (2009).

  \bibitem{ref:powheg} 
  S.~Alioli, P.~Nason, C.~Oleari, and E.~Re, 
  J. High Energy Phys. 06 (2010) 043.

  \bibitem{ref:XsecST1} 
  B.~W.~Harris, E.~Laenen, L.~Phaf, Z.~Sullivan, and S.~Weinzierl, 
  Phys. Rev. D {\bf 66}, 054024 (2002).

  \bibitem{ref:XsecST2} 
  Z.~Sullivan, 
  Phys. Rev. D {\bf 70}, 114012 (2004).

  \bibitem{ref:alpgen} 
  M.~L.~Mangano, F.~Piccinini, A.~Polosa, M.~Moretti, and R.~Pittau, 
  J. High Energy Phys. 07 (2003) 001.

  \bibitem{ref:mpt-paper} 
  M.~Bentivegna, D.~Bortoletto, Q.~Liu, F.~Margaroli, and K.~Potamianos, 
  arXiv:1205.4470 (2012).

  \bibitem{ref:single-top} 
  T. Aaltonen {\sl et al.} (CDF Collaboration), 
  Phys. Rev. D {\bf 81}, 072003 (2010).

  \bibitem{ref:tprime}   
  T. Aaltonen {\sl et al.} (CDF Collaboration), 
  Phys. Rev. Lett. {\bf 107}, 191803 (2011).

  \bibitem{ref:tom-CL} 
  T. Aaltonen {\it et al.} (CDF Collaboration), 
  Phys. Rev. D {\bf 82}, 112005 (2010).

  \bibitem{ref:Lumi} 
  D.~Acosta {\it et al.}, 
  Nucl. Instrum. Methods A {\bf 494}, 57 (2002).

\end{thebibliography}
\end{document}